\documentclass[pra,superscriptaddress,noshowpacs,epsf]{revtex4}
\usepackage{amsfonts}
\usepackage{graphicx}
\usepackage{amssymb}
\usepackage{comment}
\usepackage{amsmath}
\usepackage{color}

\begin{document}

\title{Constructions of Dicke states in high spin multi-particle systems}
\author{Wan-Fang Liu}
\email{lwf2002251@126.com}
\affiliation{School of Electric Engineering ,Anqing Teachers College, Anqing 246011,China}
\author{Zheng-Da Hu}
\email{huyuanda1112@jiangnan.edu.cn}
\affiliation{Jiangsu Provincial Research Center of Light Industrial Optoelectronic Engineering and Technology,
School of Science, Jiangnan University, Wuxi 214122, China}

\begin{abstract}
We study the constructions of Dicke states of identical particles of spin-$1$, $3/2$ and $2$
in the number representation with given particle number $N$ and magnetic quantum number $M$.
The complete bases and corresponding coefficients in the Dicke states are given, in terms of which the Dicke
states are explicitly expressed in the number representation.
As a byproduct, a rule of how to construct all the anti-symmetric states in these high spin systems is given.
Finally, by employing the negativity as the entanglement measure,
we explore the entanglement properties for spin-$1$ cases including certain pure states of two particles and
many-particle Dicke states.
\end{abstract}

\maketitle

\section{Introduction}

\label{sec:introduction}

The Dicke state, put forward by Dicke in $1954$, is a multi-particle state
of spin-$1/2$ with the maximal total angular momentum~\cite{Dicke1}. During
the past decades, it is under extensive researches and some new features
have been found. Especially, it has become a basic state as the development
of quantum information science. Based on the Dicke states, one can construct
several new quantum states such as GHZ states, W states, squeezed spin
states and spin coherence states, which are very important in quantum
information theory~\cite{G2,Du3,Ck,Wang5,Ki6}. The original Dicke state
focuses on the spin-$1/2$ case. However, the situations of spin $s$ or
angular momentum $j$ more than $1/2$ have emerged their importance and
attracted much attention as the development of low temperature physics. The
system of many $^{23}Na$ atoms trapped in a optical lattice is spin-$1$~\cite%
{Yi}, and the system of many $^{132}C_{s}$ or $^{135}B_{a}$ atoms is spin-$%
3/2$ \cite{St, Ho, Zhou}. For these high-spin systems, there may exist some
hidden symmetries, strong quantum fluctuations and novel phases~\cite{Wu}.
For instance, Haldane predicted that the one dimensional Heisenberg chain
has a spin gap for integer value of spin~\cite{Hal1, Hal2}. Wang \textit{et al}. studied the entanglement properties in a spin-$1$ Heisenberg chain~\cite%
{Wang7,Wang8}. The eigenstates and magnetic response in spin-$1$ and $2$
Bose-Einstein condensates were discussed by Koashi~\cite{Koa}. These systems
with high spin have more spin orientations and more quantum eigenstates such
that richer physical phenomena can emerge. Therefore, it is desirable to
construct the basic quantum states based on Dicke states for the high-spin
cases.

It is well known that the single particle is a qubit with only two magnetic
components for a spin-$1/2$ many-body system. Thus, the configuration of the
Dicke state $|J,M\rangle$ for spin-$1/2$ is simplest. For certain given
total spin $J$ and total magnetic component $M$, one can find the explicit
form for the its Dicke state by means of a binary linear equation group
accompanied with normalization and symmetry constraints. However, for
high-spin many-body systems, the construction of the Dicke state is not a
easy task due to much more spin components. The conventional approach of $3j$
symbol in quantum mechanics is appropriate only for the case of small
particle number $N$ and one should seek a different route for the case of
large particle number $N$. In a word, the investigation of Dicke states for
high-spin cases is a nontrivial but troublesome task which may be considered
as a supplement to the modern quantum mechanics. Motivated by the
construction of spin-$1/2$ Dicke state, in this work, we explore the Dicke
states of identical spin-$s$ particles with $s=1$, $3/2$ and $2$.

This paper is ornanized as follows. In Sec.~\ref{sec:concl}, we find those
complete bases and corresponding coefficients in the Dicke state $%
|J,M\rangle $. In Sec.~III, we study the anti-symmetric states in high spin
systems. In Sec.~IV, in terms of the negativity, the entanglement of two spin-%
$1$ particle is discussed. We conclude in Sec.~V.

\section{Construction of Dicke states in high spin multi-particle system}

\label{sec: C of Dicke}

\subsection{Dicke states in spin-1/2 multi-particle system}

First, let us recall the derivation process of Dicke states in spin-$1/2$
multi-particle system~\cite{Dicke1}. For a multi-particle system consisting
of identical spin-$1/2$ particles, the states $|J,M\rangle $, which possess
the maximal total spin angular momentum, are termed as the Dicke states.
Obviously, the states $|J,J\rangle $ and $|J,-J\rangle $ possess the simple
form
\begin{align}
|J,J\rangle =& \underset{N}{\underbrace{|\frac{1}{2}\rangle \cdots \otimes
\cdots |\frac{1}{2}\rangle }},  \notag \\
|J,-J\rangle =& \underset{N}{\underbrace{|-\frac{1}{2}\rangle \cdots \otimes
\cdots |-\frac{1}{2}\rangle }}.
\end{align}%
By introducing the collective raising and lowering operators $J_{\pm }=\sum
s_{i\pm }$ and using the relation
\begin{eqnarray}
J_{\pm }|J,M\rangle &=&\sqrt{(J\mp M)(J\pm M+1)}|J,M\pm 1\rangle ,  \notag \\
s_{i\pm }|s,m_{s}\rangle &=&\sqrt{(s\mp m_{s})(s\pm m_{s}+1)}|s,m_{s}\pm
1\rangle ,
\end{eqnarray}%
one can obtain all the other Dicke states $|J,M\rangle $ from the states $%
|J,\pm J\rangle $ although the repeating process may be tedious. Here, $s $
is the spin of single particle, $s_{i\pm }$ are raising and lowering
operators for the $i$th particle, and $m_{s}$ is the eigenvalue of $s_{z}$.
However, this method does not have any advantage for multi-qubit system,
especially, for the high-spin systems. Thereby, it is desirable to find
other ways to express Dicke states of multi-particle systems. In what
follows, we will demonstrate an alternative approach via the number
representation.

Supposing the eigenstates of the operator $s_{z}$ of spin-$1/2$ angular
momentum are $|\frac{1}{2}\rangle $ and $|-\frac{1}{2}\rangle $ and the
numbers of particles occupying the two states are $n_{1}$ and $n_{2}$,
respectively, we employ the number representation $\{|n_{1},n_{2}\rangle \}$
with
\begin{equation}
|n_{1},n_{2}\rangle =\sqrt{\frac{n_{1}!n_{2}!}{N!}}\sum P(\underset{n_{1}}{|%
\underbrace{\frac{1}{2}\cdots \frac{1}{2}}}\underset{n_{2}}{\underbrace{-%
\frac{1}{2}\cdots -\frac{1}{2}}}\rangle ),  \label{eq2}
\end{equation}%
where $N=n_{1}+n_{2}$\ is the total particle number and $P$ denotes
permutation operations between two particles with different states. Then,
the Dicke state $|J,M\rangle $ can be expressed as
\begin{equation}
|J,M\rangle =|n_{1},n_{2}\rangle ,  \label{eq1}
\end{equation}%
where $J=\frac{N}{2}$ is the maximal azimuthal quantum number of $S=\sum
s_{i}$, $M$ is the spin magnetic quantum number of $S_{z}$ whose value can
be $M=J$, $J-1$, $\cdots $, $1-J$, $-J$. It is straightforward to obtain
\begin{equation}
|J,\frac{N}{2}\rangle =|N,0\rangle ,\text{ }|J,-\frac{N}{2}\rangle
=|0,N\rangle .  \label{eq3}
\end{equation}%
In terms of Eq.~(\ref{eq2}) and the conservation of quantum numbers in
different single particle states, one can easily obtain the constraint
equation set
\begin{equation}
n_{1}+n_{2}=N, \\
\frac{n_{1}}{2}-\frac{n_{2}}{2}=M.  \label{eq4}
\end{equation}%
Therefore, once $N$ and $M$\ are given, the explicit form of the Dicke state
can be easily derived according to Eq.~(\ref{eq4}) and Eq.~(\ref{eq2}).
Finally, the Dicke state has a form

\begin{equation}
|J,M\rangle =|n_{1},n_{2}\rangle =\sqrt{\frac{(\frac{N}{2}+M)!(\frac{N}{2}%
-M)!}{N!}}\sum P(\underset{\frac{N}{2}+M}{|\underbrace{\frac{1}{2}\cdots
\frac{1}{2}}}\underset{\frac{N}{2}-M}{\underbrace{-\frac{1}{2}\cdots -\frac{1%
}{2}}}\rangle ).
\end{equation}

\label{sec: C of Dicke copy(1)}

\subsection{\protect\bigskip Dicke states of identical spin-1 particles}

\bigskip For simplicity, the states with maximal total angular momentum for $%
N$ identical particles of high spin ( $s_{i}>1/2$) are called generalized
Dicke states $|J,M\rangle $ here. Following the approach for the above spin-$%
1/2$ case, we can express $|J,M\rangle $ in the number representation as

\begin{equation}
|J,M\rangle =\sum_{k=0}^{\max
}C_{k,n_{1},n_{0},n_{-1}}|n_{1},n_{0},n_{-1}\rangle ,  \label{eq5}
\end{equation}%
where $k$ is a parameter directly related with $n_{0}$,
\begin{align}
\max =& \frac{1}{2}(J-|M|-\min ),  \notag \\
\min =& \frac{1}{2}[(-1)^{J-|M|+1}+1],  \label{eq6}
\end{align}%
and $C_{k,n_{1},n_{0},n_{-1}}$ are the superposition coefficients with $%
n_{1},$ $n_{0}$ and $n_{-1}$ denoting the numbers of particles in states $%
|\uparrow \rangle $, $|0\rangle $ and $|\downarrow \rangle $, respectively,
and
\begin{equation}
|n_{1},n_{0},n_{-1}\rangle =\sqrt{\frac{n_{1}!n_{0}!n_{-1}!}{N!}}\underset{p}%
{\sum }P(\underset{n_{1}}{|\underbrace{1\cdots 1}}\underset{n_{0}}{%
\underbrace{0\cdots 0}}\underset{n_{-1}}{\underbrace{-1\cdots -1}}\rangle ).
\label{eq7}
\end{equation}

It should be noted that the form of Eq.~(\ref{eq5}) is different from that
of Eq.~(\ref{eq1}) such that the values of $n_{1},$ $n_{0}$ and $n_{-1}$ are
not unique for specific $N$ and $M$, and they shall be determined with the
help of Eq.~(\ref{eq6}). This is the very difference for Dicke states of
high-spin systems from those of the spin-$1/2$ system. For the special cases
$M=\pm J$ and $M=\pm (J-1)$, it is easy to check that $\max =\min =0$ and $%
\max =0,$ $\min =1$, respectively, in which case $n_{1},$ $n_{0}$ and $%
n_{-1} $ are uniquely determined. For other cases, we should find all the
values of $n_{1},$ $n_{0}$ and $n_{-1}$ as well as $C_{k,n_{1},n_{0},n_{-1}}$%
.

Generally, three equations are needed to determine $n_{1},$ $n_{0}$ and $%
n_{-1}$, and it is not difficult to find the first and the second equations
as follows
\begin{align}
n_{1}+n_{0}+n_{-1}& =N,  \notag \\
n_{1}-n_{-1}& =M.  \label{eq8}
\end{align}%
Here, we find the third equation given by
\begin{equation}
n_{0}=\min +2k,  \label{eq9}
\end{equation}%
with $k=0,1,\cdots \max $. According to Eq.~(\ref{eq6}) and Eq.~(\ref{eq9}),
the number of elementary states contained in the basis $%
\{|n_{1},n_{0},n_{-1}\rangle \}$\ for state $|J,M\rangle $\ is $\max +1$.
Using the normalization condition, we also obtain the coefficients
\begin{equation}
C_{k,n_{1},n_{0},n_{-1}}=\frac{(J-|M|)!}{2^{-n_{0}}}\sqrt{\frac{N!}{%
n_{1}!n_{0}!n_{-1}!}}\overset{J-|M|}{\underset{l=1}{\prod }}\frac{1}{\sqrt{%
(2N-l+1)l}}.  \label{eq10}
\end{equation}%
It is worth nothing that an arbitrary combination of max$+1$\ elementary
states does not change the value of $M$. As a result, the arbitrary
combination also applies to the construction of the Dicke state $|J,M\rangle
$. However, since $|J,M-1\rangle $ and $|J,M\rangle $ must satisfy the
relation
\begin{equation}
|J,M-1\rangle =\frac{\hat{J}_{-}|J,M\rangle }{\sqrt{(J+M)(J-M+1)}},
\label{eq11}
\end{equation}%
the basis of $|J,M\rangle $ and that of $|J,M-1\rangle $ also satisfy
certain relations. Thereby, once the basis of the Dicke state $|J,M\rangle $
is determined, that of $|J,M-1\rangle $ is specified as well.

For an illustration, we consider the case of particle number $N=10$\ and
obtain all $\{|n_{1},n_{0},n_{-1}\rangle \}$ \ and the corresponding $%
C_{k,n_{1},n_{0},n_{-1}}$ respect to different values of $M$, which is
listed in the following Table~\ref{T1} and \ref{T22}. For convenience, we
have omitted the subscripts $k,n_{1},n_{0},n_{-1}$ in the coefficients $%
C_{k,n_{1},n_{0},n_{-1}}.$
\begin{table}[tbh]
$%
\begin{tabular}{|l|l|l|l|l|l|l|l|l|l|}
\hline
\multicolumn{2}{|l|}{$\ M=9$} & \multicolumn{2}{|l|}{$\ \ \ M=8$} &
\multicolumn{2}{|l|}{$\ \ \ M=7$} & \multicolumn{2}{|l|}{$\ \ \ M=6$} &
\multicolumn{2}{|l|}{$\ \ \ M=5$} \\ \hline
$C$ & $n_{1},n_{0},n_{-1}$ & $C$ & $n_{1},n_{0},n_{-1}$ & $C$ & $%
n_{1},n_{0},n_{-1}$ & $C$ & $n_{1},n_{0},n_{-1}$ & $C$ & $n_{1},n_{0},n_{-1}$
\\ \hline
1 & 9, 1, 0 & 0.2294 & 9, 0, 1 & 0.3794 & 8, 1, 1 & 0.0964 & 8, 0, 2 & 0.2155
& 7, 1, 2 \\ \hline
&  & 0.9733 & 8, 2, 0 & 0.9177 & 7, 3, 0 & 0.5452 & 7, 2, 1 & 0.6584 & 6, 3,
1 \\ \hline
&  &  &  &  &  & 0.8328 & 6, 4, 0 & 0.7212 & 5, 5, 0 \\ \hline
\end{tabular}%
$%
\caption{The superposition coefficient and the combination $(n_{1},$ $n_{0},$
$n_{-1})$ for every $M\in \left[ 5,9\right] $ in the spin-1 system. We set $%
N=10.$}
\label{T1}
\end{table}

\begin{table}[tbh]
$%
\begin{tabular}{|l|l|l|l|l|l|l|l|l|l|}
\hline
\multicolumn{2}{|l|}{$\ \ \ M=4$} & \multicolumn{2}{|l|}{$\ \ \ M=3$} &
\multicolumn{2}{|l|}{$\ \ \ M=2$} & \multicolumn{2}{|l|}{$\ \ \ M=1$} &
\multicolumn{2}{|l|}{$\ \ \ M=0$} \\ \hline
$C$ & $n_{1},n_{0},n_{-1}$ & $C$ & $n_{1},n_{0},n_{-1}$ & $C$ & $%
n_{1},n_{0},n_{-1}$ & $C$ & $n_{1},n_{0},n_{-1}$ & $C$ & $n_{1},n_{0},n_{-1}$
\\ \hline
0.0556 & 7, 0, 3 & 0.1472 & 6, 1, 3 & 0.0408 & 6, 0, 4 & 0.1225 & 5, 1, 4 &
0.0369 & 5, 0, 5 \\ \hline
0.3606 & 6, 2, 2 & 0.5100 & 5, 3, 2 & 0.2829 & 5, 2, 3 & 0.4473 & 4, 3, 3 &
0.2611 & 4, 2, 4 \\ \hline
0.7212 & 5, 4, 1 & 0.7212 & 4, 5, 1 & 0.6325 & 4, 4, 2 & 0.6929 & 3, 5, 2 &
0.6031 & 3, 4, 3 \\ \hline
0.5889 & 4, 6, 0 & 0.4451 & 3, 7, 0 & 0.6533 & 3, 6, 1 & 0.5238 & 2, 7, 1 &
0.6607 & 2, 6, 2 \\ \hline
&  &  &  & 0.3024 & 2, 8, 0 & 0.1746 & 1, 9, 0 & 0.3531 & 1, 8, 1 \\ \hline
&  &  &  &  &  &  &  & 0.0744 & 0, 10, 0 \\ \hline
\end{tabular}%
$%
\caption{The superposition coefficient and the combination $(n_{1},$ $n_{0},$
$n_{-1})$ for every $M\in \left[ 0,4\right] $ in the spin-1 system. We set $%
N=10.$}
\label{T22}
\end{table}
For the case of negative $M$, we need only to exchange the values between $%
n_{1}$\ and $n_{-1}$. For instance, when $J=10$, $M=-1$, in terms of the
above table, we can conveniently construct the Dicke state $|10,-1\rangle $
as
\begin{equation}
|10,-1\rangle =0.1225|4,1,5\rangle +0.4473|3,3,4\rangle +0.6929|2,5,3\rangle
+0.5238|1,7,2\rangle +0.1746|0,9,1\rangle .  \label{eq11'}
\end{equation}

\label{sec: C of Dicke copy(2)}

\subsection{Dicke states of identical spin-$3/2$ particles}

For the case of spin-$3/2$, the states have the form similar to the case of
spin-$1$, which reads
\begin{equation}
|J,M\rangle =\sum_{k=k_{0}}^{\max
}C_{k,n_{1},n_{2},n_{3},n_{4}}|n_{1},n_{2},n_{3},n_{4}\rangle ,  \label{eq12}
\end{equation}%
where $C_{k,n_{1},n_{2},n_{3},n_{4}}$ are the coefficients with $n_{1}$, $%
n_{2}$, $n_{3}$ and $n_{4}$ denoting the occupation numbers of particles in
states $|\frac{3}{2}\rangle $, $|\frac{1}{2}\rangle $, $|-\frac{1}{2}\rangle
$and $|-\frac{3}{2}\rangle $ respectively, and $k$ related to $n_{2}$ and $%
n_{3}$. The basis $\{|n_{1},n_{2},n_{3},n_{4}\rangle \}$ has the form as

\begin{equation}
|n_{1},n_{2},n_{3},n_{4}\rangle =\sqrt{\frac{n_{1}!n_{2}!n_{3}!n_{4}!}{N!}}%
\underset{P}{\sum }P(\underset{n_{1}}{|\underbrace{\frac{3}{2}\cdots \frac{3%
}{2}\rangle }}\underset{n2}{|\underbrace{\frac{1}{2}\cdots \frac{1}{2}%
\rangle }}\underset{n3}{|\underbrace{-\frac{1}{2}\cdots -\frac{1}{2}\rangle }%
}\underset{n_{4}}{|\underbrace{-\frac{3}{2}\cdots -\frac{3}{2}}}\rangle ).
\label{eq15}
\end{equation}%
With some calculations, we also derive
\begin{equation}
k_{0}=\frac{1}{2}\{\frac{1}{2}(\alpha _{1}+|\alpha _{1}|)+\frac{1}{2}%
[1-(-1)^{\frac{1}{2}(\alpha _{1}+|\alpha _{1}|)}]\},  \label{eq17}
\end{equation}%
with $\alpha _{1}=\frac{N}{2}-|M|$, and
\begin{align}
\max =& \frac{1}{2}(J-|M|-\min ),  \notag \\
\min =& \frac{1}{2}[(-1)^{J-|M|+1}+1].  \label{eq16}
\end{align}%
First, the two basic constraint equations is as follows
\begin{align}
n_{1}+n_{2}+n_{3}+n_{4}=& N,  \notag \\
3n_{1}+n_{2}-n_{3}-3n_{4}=& 2M.  \label{eq18}
\end{align}%
The third constraint equation is found to be
\begin{equation}
n_{2}-n_{3}=(-1)^{\frac{|M|-M}{2|M|}}\gamma _{_{1}},  \label{eq19}
\end{equation}%
with $\gamma _{_{1}}=J-|M|-3k$. For specific value of $k$, we also find the
fourth constraint equation as
\begin{equation}
n_{2}+n_{3}=|\gamma _{_{1}}|-2(k_{1}-1),  \label{eq20}
\end{equation}%
with
\begin{equation}
m_{k}=\frac{1}{2}[\frac{1}{2}(\beta _{1}-|\beta _{1}|)+k+1+|\frac{1}{2}%
(\beta _{1}-|\beta _{1}|)+k+1|]+\frac{1}{2}(\gamma _{1}+|\gamma _{1}|)
\label{eq21}
\end{equation}%
and $\beta _{1}=k-\frac{1}{2}(\alpha _{1}+|\alpha _{1}|)$. Then, we obtain a
set of constraint equations
\begin{equation}
\underset{k=k_{0}}{\overset{\max }{\sum }}\overset{m_{k}}{\underset{k_{1}=1}{%
\sum }}\left\{
\begin{array}{c}
n_{1}+n_{2}+n_{3}+n_{4}=N \\
3n_{1}+n_{2}-n_{3}-3n_{4}=2M \\
n_{2}-n_{3}=(-1)^{\frac{|M|-M}{2|M|}}\gamma _{_{1}} \\
n_{2}+n_{3}=|\gamma _{_{1}}|-2(k_{1}-1)%
\end{array}%
\right.  \label{eq22}
\end{equation}%
with $k_{0}\leq k\leq \max $ and $1\leq k_{1}\leq m_{k}$. The number of
states to form a complete basis is $\underset{k=k_{0}}{\overset{\max }{\sum }%
}m_{k}$\ and the normalized coefficients are
\begin{equation}
C_{k,n_{1},n_{2},n_{3},n_{4}}=\frac{(J-|M|)!}{3^{-(n_{2}+n_{3})/2}}\sqrt{%
\frac{N!}{n_{1}!n_{2}!n_{3}!n_{4!}}}\overset{J-|M|}{\underset{l=1}{\prod }}%
\frac{1}{\sqrt{(3N-l+1)l}}.  \label{eq23}
\end{equation}%
Again, for an illustration, we consider the case of particle number $N=6$,
and list all these parameters in the Dicke states for different values of $M$
in Table~\ref{T33} and \ref{T44}
\begin{table}[tbh]
$%
\begin{tabular}{|l|l|l|l|l|l|l|l|l|l|}
\hline
\multicolumn{2}{|l|}{$\ \ M=9$} & \multicolumn{2}{|l|}{$\ \ M=8$} &
\multicolumn{2}{|l|}{$\ \ \ M=7$} & \multicolumn{2}{|l|}{$\ \ \ \ M=6$} &
\multicolumn{2}{|l|}{$\ \ \ \ M=5$} \\ \hline
$C$ & $n_{1},n_{2},n_{3},n_{4}$ & $C$ & $n_{1},n_{2},n_{3},n_{4}$ & $C$ & $%
n_{1},n_{2},n_{3},n_{4}$ & $C$ & $n_{1},n_{2},n_{3},n_{4}$ & $C$ & $%
n_{1},n_{2},n_{3},n_{4}$ \\ \hline
1 & 6, 0, 0, 0 & 1 & 5, 1, 0, 0 & 0.9393 & 4, 2, 0, 0 & 0.8135 & 3, 3, 0, 0
& 0.6301 & 2, 4, 0, 0 \\ \hline
&  &  &  & 0.3430 & 5, 0, 1, 0 & 0.0857 & 5, 0, 0, 1 & 0.1715 & 4, 1, 0, 1
\\ \hline
&  &  &  &  &  & 0.5752 & 4,1,1,0 & 0.2100 & 4,0,2,0 \\ \hline
&  &  &  &  &  &  &  & 0.7276 & 3, 2, 1, 0 \\ \hline
\end{tabular}%
$%
\caption{The superposition coefficient and the combination $(n_{1},$ $n_{2},$
$n_{3},$ $n_{4})$ for every $M\in \left[ 5,9\right] $ in the spin-3/2
system. We set $N=6.$}
\label{T33}
\end{table}
\begin{table}[tbh]
$%
\begin{tabular}{|l|l|l|l|l|l|l|l|l|l|}
\hline
\multicolumn{2}{|l|}{$\ \ \ \ M=4$} & \multicolumn{2}{|l|}{$\ \ \ \ M=3$} &
\multicolumn{2}{|l|}{$\ \ \ \ M=2$} & \multicolumn{2}{|l|}{$\ \ \ \ M=1$} &
\multicolumn{2}{|l|}{$\ \ \ \ M=0$} \\ \hline
$C$ & $n_{1},n_{2},n_{3},n_{4}$ & $C$ & $n_{1},n_{2},n_{3},n_{4}$ & $C$ & $%
n_{1},n_{2},n_{3},n_{4}$ & $C$ & $n_{1},n_{2},n_{3},n_{4}$ & $C$ & $%
n_{1},n_{2},n_{3},n_{4}$ \\ \hline
0.4125 & 1, 5, 0, 0 & 0.1982 & 0, 6, 0, 0 & 0.2763 & 1, 4, 0, 1 & 0.1825 &
0, 5, 0, 1 & 0.1825 & 1, 3, 0, 2 \\ \hline
0.2510 & 3, 2, 0, 1 & 0.2954 & 2, 3, 0, 1 & 0.0752 & 3, 1, 0, 2 & 0.1361 &
2, 2, 0, 2 & 0.0203 & 3, 0, 0, 3 \\ \hline
0.1025 & 4, 0, 1, 1 & 0.0284 & 4, 0, 0, 2 & 0.1303 & 3, 0, 2, 1 & 0.0641 &
3, 0, 1, 2 & 0.3872 & 0, 4, 1, 1 \\ \hline
0.7531 & 2, 3, 1, 0 & 0.1706 & 3, 0, 3, 0 & 0.3707 & 0, 5, 1, 0 & 0.1666 &
2, 0, 4, 0 & 0.1825 & 2, 0, 3, 1 \\ \hline
0.4348 & 3, 1, 2, 0 & 0.6267 & 1, 4, 1, 0 & 0.3908 & 2, 2, 1, 1 & 0.4713 &
1, 3, 1, 1 & 0.1825 & 2, 1, 1, 2 \\ \hline
&  & 0.2412 & 3, 1, 1, 1 & 0.3908 & 2, 1, 3, 0 & 0.3333 & 2, 1, 2, 1 & 0.3872
& 1, 1, 4, 0 \\ \hline
&  & 0.6267 & 2, 2, 2, 0 & 0.6769 & 1, 3, 2, 0 & 0.4999 & 0, 4, 2, 0 & 0.5476
& 1, 2, 2, 1 \\ \hline
&  &  &  &  &  & 0.5772 & 1, 2, 3, 0 & 0.5476 & 0, 3, 3, 0 \\ \hline
\end{tabular}%
$%
\caption{The superposition coefficient and the combination $(n_{1},$ $n_{2},$
$n_{3},$ $n_{4})$ for every $M\in \left[ 0,4\right] $ in the spin-3/2
system. We set $N=6.$}
\label{T44}
\end{table}

For negative values of $M$, one may perform the exchanges $%
n_{1}\rightleftarrows n_{4}$, $n_{2}\rightleftarrows n_{3}$ in the case of
positive $M$. For instance, when $J=6$, $M=-1$, using the results of above
table, we obtain

\begin{align}
|6,-1\rangle =& 0.1825|1,0,5,0\rangle +0.1361|2,0,2,2\rangle
+0.0641|2,1,0,3\rangle +0.1666|0,4,0,2\rangle +0.4713|1,1,3,1\rangle  \notag
\\
\quad&+0.3333|1,2,1,2\rangle +0.4999|0,2,4,0\rangle +0.5772|0,3,2,1\rangle.
\label{eq24}
\end{align}

\label{sec: C of Dicke copy(3)}

\subsection{Dicke states of identical spin-2 particles}

For the case of spin-$2$, the Dicke states in the number representation are
given by

\begin{equation}
|J,M\rangle =\sum_{k=k_{0}}^{\max
}C_{k,n_{1},n_{2},n_{3},n_{4},n_{5}}|n_{1},n_{2},n_{3},n_{4},n_{5}\rangle ,
\label{eq25}
\end{equation}%
with $J=2N$ and the number states
\begin{equation}
|n_{1},n_{2},n_{3},n_{4},n_{5}\rangle =\sqrt{\frac{%
n_{1}!n_{2}!n_{3}!n_{4}!n_{5}!}{N!}}\underset{P}{\sum }P(\underset{n_{1}}{|%
\underbrace{2\cdots 2\rangle }}\underset{n2}{\underbrace{|1\cdots 1\rangle }}%
\underset{n_{3}}{\underbrace{|0\cdots 0}}\underset{n4}{\underbrace{|-1\cdots
-1\rangle }}\underset{n5}{\underbrace{|-2\cdots -2}}\rangle ),  \label{eq26}
\end{equation}%
where $n_{1}$, $n_{2}$, $n_{3}$, $n_{4}$ and $n_{5}$ denote the numbers of
particles in states $|2\rangle $, $|1\rangle $, $|0\rangle $, $|-1\rangle $
and $|-2\rangle $ respectively, $k$ is related to $n_{2}$ and $n_{4}$, and $%
k_{0}$ and $\max $ are given by
\begin{equation}
k_{0}=\left\{
\begin{array}{c}
\frac{1}{3}(\alpha _{2}+|\alpha _{2}|),\text{ if }\frac{1}{2}(\alpha
_{2}+|\alpha _{2}|)=3n^{\prime }, \\
\frac{1}{3}(\alpha _{2}+|\alpha _{2}|+1),\text{ if }\frac{1}{2}(\alpha
_{2}+|\alpha _{2}|)=3n^{\prime }+1, \\
\frac{1}{3}(\alpha _{2}+|\alpha _{2}|+2),\text{ if }\frac{1}{2}(\alpha
_{2}+|\alpha _{2}|)=3n^{\prime }+2,%
\end{array}%
\right.  \label{eq27}
\end{equation}%
\begin{equation}
\max =\left\{
\begin{array}{c}
\frac{2}{3}(2N-|M|),\text{ if }2N-|M|=3n^{\prime \prime }, \\
\frac{2}{3}(2N-|M|-\frac{1}{2}),\text{ if }2N-|M|=3n^{\prime \prime }-1, \\
\frac{2}{3}(2N-|M|-1),\text{ if }2N-|M|=3n^{\prime \prime }-2.%
\end{array}%
\right.  \label{eq28}
\end{equation}%
Here, $n^{\prime }$ and $n^{\prime \prime }$ are integers and $\alpha
_{2}=N-|M|$. It is easy to verify that
\begin{equation}
|J,J\rangle =|N,0,0,0,0\rangle ,  \notag \\
|J,-J\rangle =|0,0,0,0,N\rangle ,  \label{eq29}
\end{equation}%
To specify $n_{1}$, $n_{2}$, $n_{3}$, $n_{4}$ and $n_{5}$ as well as $%
C_{k,n_{1},n_{2},n_{3},n_{4},n_{5}}$, five equations are needed. First, the
two basic equations are
\begin{align}
n_{1}+n_{2}+n_{3}+n_{4}+n5=& N,  \notag \\
2n_{1}+n_{2}-n_{4}-2n_{5}=& 2M.  \label{eq30}
\end{align}%
We derive the other three equations as
\begin{align}
&n_{2}-n_{4}= (-1)^{\frac{|M|-M}{2|M|}}\gamma _{_{2}},\text{ }k_{0}\leq
k\leq \max ,\quad \gamma _{_{2}}=J-|M|-2k,  \notag  \label{eq31} \\
&n_{2}+n_{4}= \gamma _{_{2}}+2(k_{1}-1),\quad 1\leq k_{1}\leq m_{k},  \notag
\\
&n_{3}= 2(k_{2}+1)+\frac{1}{2}[1-(-1)^{k}]-2,\quad 0\leq k_{2}\leq
m_{k}-k_{1},
\end{align}%
where
\begin{equation}
m_{k}=\frac{1}{2}\{\frac{1}{2}(\beta _{2}+|\beta _{2}|)+\frac{2k+3+(-1)^{k}}{%
4}+|\frac{1}{2}(\beta _{2}+|\beta _{2}|)+\frac{2k+3+(-1)^{k}}{4}|\}+\frac{1}{%
2}(\gamma _{2}-|\gamma _{2}|),  \label{eq33}
\end{equation}%
with $\beta _{2}=k-\frac{1}{2}(\alpha _{2}+|\alpha _{2}|)$. Note that $k$, $%
k_{1}$ and $k_{2}$ are all nonnegative integers. Besides, we find that the
number of states to form a complete basis is $\overset{\max }{\underset{%
k=k_{0}}{\sum }}\frac{1}{2}m_{k}(m_{k}+1)$ and the normalized coefficients
read as
\begin{equation}
C_{k,n_{1},n_{2},n_{3},n_{4},n_{5}}=(\frac{3}{2})^{n_{3}/2}\frac{(J-|M|)!}{%
3^{-(n_{2}+n_{3}+n_{4})/2}}\sqrt{\frac{N!}{n_{1}!n_{2}!n_{3}!n_{4!}!n_{5}}}%
\overset{J-|M|}{\underset{l=1}{\prod }}\frac{1}{\sqrt{(4N-l+1)l}}.
\label{eq34}
\end{equation}

Here, we focus on the case $N=5$ for example and derive all the ($n_{1},$ $%
n_{2},$ $n_{3},$ $n_{4},$ $n_{5})$ and $C_{k,n_{1},n_{2},n_{3},n_{4},n_{5}}$
for different values of $M$\ as follows. In terms of the table above, all
the five-particle Dicke states can be obtained.
\begin{table}[tbh]
$%
\begin{tabular}{|l|l|l|l|l|l|l|l|l|l|}
\hline
\multicolumn{2}{|l|}{$\ \ \ \ M=9$} & \multicolumn{2}{|l|}{$\ \ \ \ \ \ M=8$}
& \multicolumn{2}{|l|}{$\ \ \ \ \ \ \ \ M=7$} & \multicolumn{2}{|l|}{$\ \ \
\ \ \ \ M=6$} & \multicolumn{2}{|l|}{$\ \ \ \ \ \ \ \ M=5$} \\ \hline
$C$ & $n_{1},n_{2},n_{3},n_{4},n_{5}$ & $C$ & $n_{1},n_{2},n_{3},n_{4},n_{5}$
& $C$ & $n_{1},n_{2},n_{3},n_{4},n_{5}$ & $C$ & $%
n_{1},n_{2},n_{3},n_{4},n_{5}$ & $C$ & $n_{1},n_{2},n_{3},n_{4},n_{5}$ \\
\hline
1 & 4, 1, 0, 0, 0 & 0.9177 & 3, 2, 0, 0, 0 & 0.7493 & 2, 3, 0, 0, 0 & 0.5140
& 1, 4, 0, 0, 0 & 0.2570 & 0, 5, 0, 0, 0 \\ \hline
&  & 0.3974 & 4, 0, 1, 0, 0 & 0.6489 & 3, 1, 1, 0, 0 & 0.7710 & 2, 2, 1, 0, 0
& 0.7038 & 1, 3, 1, 0, 0 \\ \hline
&  &  &  & 0.1325 & 4, 0, 0, 1, 0 & 0.0321 & 4, 0, 0, 0, 1 & 0.0718 & 3, 1,
0, 0, 1 \\ \hline
&  &  &  &  &  & 0.2726 & 3, 0, 2, 0, 0 & 0.5279 & 2, 1, 2, 0, 0 \\ \hline
&  &  &  &  &  & 0.2570 & 3, 1, 0, 1, 0 & 0.3519 & 2, 2, 0, 1, 0 \\ \hline
&  &  &  &  &  &  &  & 0.1760 & 3, 1, 0, 1, 0 \\ \hline
\end{tabular}%
$%
\caption{The superposition coefficient and the combination $(n_{1},$ $n_{2},$
$n_{3},$ $n_{4},$ $n_{5})$ for every $M\in \left[ 5,9\right] $ in the spin-2
system. We set $N=5.$}
\label{T5}
\end{table}
\begin{table}[tbh]
$%
\begin{tabular}{|l|l|l|l|l|l|l|l|l|l|}
\hline
\multicolumn{2}{|l|}{$\ \ \ \ M=4$} & \multicolumn{2}{|l|}{$\ \ \ \ \ \ M=3$}
& \multicolumn{2}{|l|}{$\ \ \ \ \ \ \ \ M=2$} & \multicolumn{2}{|l|}{$\ \ \
\ \ \ \ M=1$} & \multicolumn{2}{|l|}{$\ \ \ \ \ \ \ \ M=0$} \\ \hline
$C$ & $n_{1},n_{2},n_{3},n_{4},n_{5}$ & $C$ & $n_{1},n_{2},n_{3},n_{4},n_{5}$
& $C$ & $n_{1},n_{2},n_{3},n_{4},n_{5}$ & $C$ & $%
n_{1},n_{2},n_{3},n_{4},n_{5}$ & $C$ & $n_{1},n_{2},n_{3},n_{4},n_{5}$ \\
\hline
0.1113 & 0, 4, 1, 0, 0 & 0.1285 & 1, 3, 0, 0, 1 & 0.1008 & 0, 4, 0, 0, 1 &
0.2138 & 0, 3, 1, 0, 1 & 0.0510 & 1, 2, 0, 0, 2 \\ \hline
0.6677 & 2, 2, 0, 0, 1 & 0.5452 & 0, 3, 2, 0, 0 & 0.2138 & 1, 2, 1, 0, 1 &
0.0267 & 2, 1, 0, 0, 2 & 0.3058 & 0, 2, 2, 0, 1 \\ \hline
0.3634 & 1, 2, 2, 0, 0 & 0.2570 & 0, 4, 0, 1, 0 & 0.5238 & 0, 2, 3, 0, 0 &
0.2268 & 1, 1, 2, 0, 1 & 0.1665 & 0, 3, 0, 1, 1 \\ \hline
0.0556 & 1, 3, 0, 1, 0 & 0.1363 & 2, 1, 1, 0, 1 & 0.4938 & 0, 3, 1, 1, 0 &
0.3928 & 0, 1, 4, 0, 0 & 0.0312 & 2, 0, 1, 0, 2 \\ \hline
0.2361 & 3, 0, 1, 0, 1 & 0.4721 & 1, 1, 3, 0, 0 & 0.0089 & 3, 0, 0, 0, 2 &
0.1512 & 1, 2, 0, 1, 1 & 0.1529 & 1, 0, 3, 0, 1 \\ \hline
0.3855 & 2, 0, 3, 0, 0 & 0.5452 & 1, 2, 1, 1, 0 & 0.0926 & 2, 0, 2, 0, 1 &
0.6415 & 0, 2, 2, 1, 0 & 0.2052 & 0, 0, 5, 0, 0 \\ \hline
0.4451 & 2, 1, 1, 1, 0 & 0.0321 & 3, 0, 0, 1, 1 & 0.2268 & 1, 0, 4, 0, 0 &
0.2469 & 0, 3, 0, 2, 0 & 0.2497 & 1, 1, 1, 1, 1 \\ \hline
0.06423 & 3, 0, 0, 2, 0 & 0.2361 & 2, 0, 2, 1, 0 & 0.1008 & 0, 4, 0, 0, 1 &
0.0926 & 2, 0, 1, 1, 1 & 0.6116 & 0, 1, 3, 1, 0 \\ \hline
&  & 0.1574 & 2, 1, 0, 2, 0 & 0.0873 & 2, 1, 0, 1, 1 & 0.3208 & 1, 0, 3, 1, 0
& 0.4994 & 0, 2, 1, 2, 0 \\ \hline
&  &  &  & 0.5238 & 1, 1, 2, 1, 0 & 0.3704 & 1, 1, 1, 2, 0 & 0.0510 & 2, 0,
0, 2, 1 \\ \hline
&  &  &  & 0.2469 & 1, 2, 0, 2, 0 & 0.0617 & 2, 0, 0, 3, 0 & 0.3058 & 1, 0,
2, 2, 0 \\ \hline
&  &  &  & 0.1512 & 2, 0, 1, 2, 0 &  &  & 0.1665 & 1, 1, 0, 3, 0 \\ \hline
\end{tabular}%
$%
\caption[table2]{The superposition coefficient and the combination $(n_{1},$
$n_{2},$ $n_{3},$ $n_{4},$ $n_{5})$ for every $M\in \left[ 0,4\right] $ in
the spin-2 system. We set $N=5.$}
\label{T6}
\end{table}
According to the expression of the superposition coefficients $C$ of spin-$1$%
, $3/2$ and $2$, it is not difficult to conclude that the coefficients for
spin-$1/2$ case can be expressed as
\begin{equation}
C_{k,n_{1},n_{2}}=\sqrt{\frac{N!}{n_{1}!n_{2}!}}(J-|M|)!\overset{J-|M|}{%
\underset{l=1}{\prod }}\frac{1}{\sqrt{(N-l+1)l}},  \label{eq35}
\end{equation}%
By solving Eq.~(\ref{eq4}) to obtain
\begin{equation}
J-|M|=\left\{
\begin{array}{c}
n_{1}\text{ if }M>0, \\
n_{2}\text{ if }M<0,%
\end{array}%
\right.  \label{eq36}
\end{equation}%
and substituting Eq.~(\ref{eq36}) into Eq.~(\ref{eq35}), the coefficients
are simplified as%
\begin{equation}
C_{k,n_{1},n_{2}}\equiv 1,  \label{eq37}
\end{equation}%
which agrees with the setup in Eq.~(\ref{eq1}).

\section{Anti-symmetric states in high spin systems}

As a natural byproduct, we proceed to discuss the anti-symmetric states in
these high spin systems. It is well known that any two particles should not
be in the same state in an anti-symmetric state. Therefore, we can conclude
that anti-symmetric states exist only for particle number less than $2s+1$
for spin-$s$ systems. For the case of electrons of spin-$1/2$ , the particle
number of collective anti-symmetric spin states is only $2$. Similarly, one
can check that the upper limits for particle numbers of spin $1$, $3/2$ and $%
2$ are $3$, $4$ and $5$, respectively. Based on this fact, the number of
anti-symmetric states for high spin-$s$ systems is
\begin{equation}
C(2s+1,2s+1)+C(2s+1,2s)+\cdots +C(2s+1,2)=2^{2s+1}-(2s+2),  \label{tas}
\end{equation}%
where $C(n,k)$ means the $k$-combinations of $n$.

For convenience, we mark the states of single particle as $|\alpha \rangle ,$
$|\beta \rangle \cdots \in \{|-s\rangle ,\cdots ,|s\rangle \}$. Then, for
the case of two particles, the $C_{2s+1}^{2}$ elementary anti-symmetric
states can be write as
\begin{equation}
|\psi (i,j)\rangle _{AS}=\sqrt{\frac{1}{2}}\underset{P}{\sum }\delta
_{P}P(|\alpha \rangle |\beta \rangle ),\text{ }|\alpha \rangle \neq |\beta
\rangle ,  \label{an}
\end{equation}%
where the symbol $P$\ denotes all possible permutations and $\delta _{P}$
(with the initial value +1) changes its sign between $+1$ and $-1$ after
every permutation. Due to the fact that the state $|J=2s-1,2s-1\rangle $ is
anti-symmetric, so all states belonging to the subspace with total spin $%
J=2s-1$ are anti-symmetric. It is obvious that in the two-qubit system there
is only a elementary anti-symmetric state

\begin{equation}
|J=0,M=0\rangle =\sqrt{\frac{1}{2}}(|\uparrow \rangle |\downarrow \rangle
-|\downarrow \rangle |\uparrow \rangle ),
\end{equation}%
and in two-qudit system with $J=1$ there is three elementary anti-symmetric
states as

\begin{eqnarray}
|J &=&1,1\rangle =\sqrt{\frac{1}{2}}(|\uparrow \rangle |0\rangle -|0\rangle
|\uparrow \rangle ),  \notag \\
|J &=&1,0\rangle =\sqrt{\frac{1}{2}}(|\uparrow \rangle |\downarrow \rangle
-|\downarrow \rangle |\uparrow \rangle ),  \notag \\
|J &=&1,-1\rangle =\sqrt{\frac{1}{2}}(|0\rangle |\downarrow \rangle
-|\downarrow \rangle |0\rangle ),
\end{eqnarray}
For the case with $s>1,$ in the subspace with total spin $J=2s-1,$ the four
states
\begin{eqnarray}
|J &=&2s-1,\pm (2s-1)\rangle =\sqrt{\frac{1}{2}}(|\pm s\rangle |\pm s\mp
1\rangle -|\pm s\mp 1\rangle |\pm s\rangle ),  \notag \\
|J &=&2s-1,\pm (2s-2)\rangle =\sqrt{\frac{1}{2}}(|\pm s\rangle |\pm s\mp
2\rangle -|\pm s\mp 2\rangle |\pm s\rangle ),
\end{eqnarray}
are elementary anti-symmetric states given by Eq.~(\ref{an}), and the rest $%
4s-5$ states are linear superposition of those elementary anti-symmetric
states. For the case including three particles, there is $C_{2s+1}^{3}$
elementary anti-symmetric states\bigskip
\begin{equation}
|\psi (i,j,k)\rangle _{AS}=\sqrt{\frac{1}{3!}}\underset{P}{\sum }\delta
_{P}P(|\alpha \rangle |\beta \rangle |\gamma \rangle ),\text{ }|\alpha
\rangle \neq |\beta \rangle \neq |\gamma \rangle .
\end{equation}

To conclude, for the $2s+1$ particles, there is only one elementary
anti-symmetric state as

\begin{equation}
|\psi \lbrack 1,2\cdots (2s+1)\text{th}]\rangle _{AS}=\sqrt{\frac{1}{(2s+1)!}%
}\underset{P}{\sum }\delta _{P}P(|-s\rangle \otimes |-s+1\rangle \cdots
\otimes \cdots |s-1\rangle \otimes |s\rangle ),
\end{equation}

which is just the state $|J=0,M=0\rangle $. In particular, for the case of
spin-1, the anti-symmetric state can be written as%
\begin{equation}
|\psi (1,2,3)\rangle _{AS}=\sqrt{\frac{1}{3!}}(|\uparrow \rangle |0\rangle
|\downarrow \rangle -|\uparrow \rangle |\downarrow \rangle |0\rangle
+|\downarrow \rangle |0\rangle |\uparrow \rangle -|\downarrow \rangle
|\uparrow \rangle |0\rangle -|0\rangle |\uparrow \rangle |\downarrow \rangle
+|0\rangle |\downarrow \rangle |\uparrow \rangle ).
\end{equation}

For instance, the particle number may be $2$, $3$, $4$ and $5$ for $s=2$,
and one can construct all the anti-symmetric states as
\begin{align}
|\psi (i,j)\rangle _{AS}=& \sqrt{\frac{1}{2!}}\underset{P}{\sum }\delta
_{P}P(|\alpha \rangle |\beta \rangle ),\text{ }|\alpha \rangle \neq |\beta
\rangle ,  \notag \\
|\psi (i,j,k)\rangle _{AS}=& \sqrt{\frac{1}{3!}}\underset{P}{\sum }\delta
_{P}P(|\alpha \rangle |\beta \rangle |\gamma \rangle ),\text{ }|\alpha
\rangle \neq |\beta \rangle \neq |\gamma \rangle ,  \notag \\
|\psi (i,j,k,l)\rangle _{AS}=& \sqrt{\frac{1}{4!}}\underset{P}{\sum }\delta
_{P}P(|\alpha \rangle |\beta \rangle |\gamma \rangle |\eta \rangle ),|\alpha
\rangle \neq |\beta \rangle \neq |\gamma \rangle \neq |\eta \rangle ,  \notag
\\
|\psi (i,j,k,l,m)\rangle _{AS}=& \sqrt{\frac{1}{5!}}\underset{P}{\sum }%
\delta _{P}P(|2\rangle |1\rangle |0\rangle |-1\rangle |-2\rangle ),
\label{eq38}
\end{align}%
where $|\alpha \rangle $, $|\beta \rangle $, $|\gamma \rangle $, $|\eta
\rangle \in $ $\{|2\rangle ,$ $|1\rangle ,$ $|0\rangle ,$ $|-1\rangle $, $%
|-2\rangle \},$ and $|2\rangle ,$ $|1\rangle ,$ $|0\rangle ,$ $|-1\rangle $
and $|-2\rangle $ are the eigenstates of single spin magnetic quantum number
$m_{s}$ with eigenvalues $2$, $1$, $0$, $-1$ and $-2$, respectively.
According Eq.~(\ref{tas}), there totally exist $26$ different anti-symmetric
states for spin-$2$ systems. Generally speaking, due to the Pauli exclusion
principle, the number of anti-symmetric states compared to that of all the
Dicke states is very limited.

\section{Entanglement of two qudits in system with many particles}

In this section, we study the entanglement for the case of two qudits. The
entanglement criteria proposed by Peres-Horodecki~\cite{P1, P2} is adopted.
For states with certain symmetries in the high spin systems, this criterion
is good enough to measure the entanglement. However, the states discussed by
us are beyond this requirement. In order to quantify the entanglement, Vidal
and Werner proposed a entanglement measure termed as negativity~\cite{Wi}.
The first thing need to do is that one obtains the density matrix $\rho
_{ij} $\ of two qudits in the basis \{$|\uparrow \downarrow \rangle ,$ $%
|00\rangle ,$ $|\downarrow \uparrow \rangle ,$ $|\uparrow \rangle |0\rangle ,
$ $|0\rangle |\uparrow \rangle ,$ $|0\rangle |\downarrow \rangle ,$ $%
|\downarrow \rangle |0\rangle |,$ $|\uparrow \rangle |\uparrow \rangle ,$ $%
|\downarrow \rangle |\downarrow \rangle $\}. Next, one can perform partial
transpose (PT) to $\rho _{ij}$, and obtain the matrix $\rho _{ij}^{T}$ in
the basis spanned by \{$|\uparrow \rangle |\uparrow \rangle ,$ $|0\rangle
|0\rangle ,$ $|\downarrow \rangle |\downarrow \rangle ,$ $|\uparrow \rangle
|0\rangle ,$ $|0\rangle |\downarrow \rangle ,$ $|0\rangle |\uparrow \rangle
, $ $|\downarrow \rangle |0\rangle ,$ $|\uparrow \rangle |\downarrow \rangle
,$ $|\downarrow \rangle |\uparrow \rangle $\}. The negativity is then
defined as
\begin{equation}
\mathcal{N}(\rho _{ij})=\underset{i}{\sum }|\lambda _{i}|,  \label{eq39}
\end{equation}%
where $\lambda _{i}$ are the negative eigenvalues of $\rho _{ij}^{T}$. If $%
\mathcal{N}(\rho _{ij})>0$, then the two particles stay in the entangled
state. However, in the $9$ basis, the density of two particles generally has
$81$ elements. Different from the case of many spin-$1/2$ particles, these
elements can not be represented by the expectation value of the collective
operators of system. This leads to certain difficulties in the calculation
of the entanglement. However, in the following, we will show that the number
of effective elements will be greatly reduced for some special states.

\subsection{Entanglement of specific states in the system with two spin-$1$
particles}

Let us begin with the system with two spin-1 particles. We discuss the
entanglement of two particle with spin-$1$. The first state considered is
the generalized symmetric Bell state of two qudits, which is an important
state for the qudit teleportation scheme~\cite{Sych,Be}. Its form is given
by
\begin{equation}
|B_{G}\rangle =\sqrt{\frac{1}{3}}(|\uparrow \rangle |\uparrow \rangle
+|0\rangle |0\rangle +|\downarrow \rangle |\downarrow \rangle ),
\label{eq40}
\end{equation}%
where $s_{z}|\uparrow \rangle =|\uparrow \rangle $, $s_{z}|0\rangle =0$, and
$s_{z}|\downarrow \rangle =|\downarrow \rangle $. It is easy to check that
this state is a maximally entangled state and the negativity equal 1 in this
state. In order to compare with the state presented above, we study the
negativity of another state with a generalized form
\begin{equation}
|\psi _{1}\rangle =\sqrt{\frac{1}{3}}[|\uparrow \rangle |\uparrow \rangle
+c_{1}\frac{1}{\sqrt{2}}(|\uparrow \rangle |\downarrow \rangle +|\downarrow
\rangle |\uparrow \rangle )+c_{2}|0\rangle |0\rangle +|\downarrow \rangle
|\downarrow \rangle ],  \label{eq41}
\end{equation}%
where coefficients $c_{1}$ and $c_{2}$ satisfy the relation $%
|c_{1}|^{2}+|c_{2}|^{2}=1$. After the PT, we can give the matrix density of
two particles in a block diagonal form as
\begin{equation}
\rho _{ij}^{T}=\mathrm{diag}(C_{5\times 5},D_{4\times 4}),  \label{eq411}
\end{equation}%
with $C$ and $D$ given by
\begin{equation}
C=\left(
\begin{array}{ccccc}
\frac{1}{3} & 0 & \frac{|c_{1}|^{2}}{6} & \frac{c_{1}^{\ast }}{3\sqrt{2}} &
\frac{c_{1}}{3\sqrt{2}} \\
0 & \frac{|c_{2}|^{2}}{3} & 0 & 0 & 0 \\
\frac{|c_{1}|^{2}}{6} & 0 & \frac{1}{3} & \frac{c_{1}}{3\sqrt{2}} & \frac{%
c_{1}^{\ast }}{3\sqrt{2}} \\
\frac{c_{1}}{3\sqrt{2}} & 0 & \frac{c_{1}^{\ast }}{3\sqrt{2}} & \frac{%
|c_{1}|^{2}}{6} & \frac{1}{3} \\
\frac{c_{1}^{\ast }}{3\sqrt{2}} & 0 & \frac{c_{1}}{3\sqrt{2}} & \frac{1}{3}
& \frac{|c_{1}|^{2}}{6}%
\end{array}%
\right) ,\quad D=\left(
\begin{array}{cccc}
0 & \frac{c_{2}c_{1}^{\ast }}{3\sqrt{2}} & \frac{c_{2}}{3} & 0 \\
\frac{c_{1}c_{2}^{\ast }}{3\sqrt{2}} & 0 & 0 & \frac{c_{2}^{\ast }}{3} \\
\frac{c_{2}^{\ast }}{3} & 0 & 0 & \frac{c_{1}c_{2}^{\ast }}{3\sqrt{2}} \\
0 & \frac{c_{2}}{3} & \frac{c_{2}c_{1}^{\ast }}{3\sqrt{2}} & 0%
\end{array}%
\right) .  \label{eq42}
\end{equation}%
Specially, for combination ($c_{1}=\sqrt{1/3}$ and $c_{2}$=$\sqrt{2/3}$),
the state can be reduced to even spin coherent state of two qudits
\begin{equation}
|\psi _{e}\rangle =\sqrt{\frac{1}{3}}(|2,2\rangle +|2,0\rangle +|2,-2\rangle
).  \label{eq43}
\end{equation}%
The negativity in this state is $0.8221$, where $|2,0\rangle $ is the Dicke
state
\begin{equation}
|2,0\rangle =\sqrt{\frac{1}{6}}(|\uparrow \rangle |\downarrow \rangle
+|\downarrow \rangle |\uparrow \rangle )+\sqrt{\frac{2}{3}}|0\rangle
|0\rangle ,  \label{eq44}
\end{equation}%
in which state the negativity is $0.833$. It is worth mentioning that this
even spin coherent state can be generate by one-axis twisting model or the
two-axis counter model with the initial state $|2,-2\rangle $. For another
state
\begin{equation}
|\psi _{2}\rangle =\frac{1}{2}(|\uparrow \rangle |\downarrow \rangle
+|\downarrow \rangle |\uparrow \rangle )+\frac{1}{\sqrt{2}}|0\rangle
|0\rangle ,  \label{eq45}
\end{equation}%
with $M=0$, we obtain the negativity with value $0.9571.$

There are two interested generalized singlet Bell states which are the
elementary states of dimmer states \cite{Sych,Chu,Ar}
\begin{equation}
|B_{s\pm }\rangle =\sqrt{\frac{1}{3}}(|\uparrow \rangle |\downarrow \rangle
+|\downarrow \rangle |\uparrow \rangle \pm |0\rangle |0\rangle ).
\label{eq46}
\end{equation}%
Also, we obtain that the negativity is $1$. \label{sec: C of Dicke copy(5)}

\subsection{\protect\bigskip Entanglement of two qudits in the Dicke states
of many spin-1 particles}

In the nine bases, the reduced density matrix of the two spins in the Dicke
states can be written as
\begin{equation}
\rho _{ij}=\mathrm{diag}(T_{1},T_{2,}\text{ }T_{3},\text{ }a_{8},\text{ }%
a_{9})  \label{eq47}
\end{equation}%
with $T_{1},T_{2}$ and $T_{3}$
\begin{equation}
T_{1}=\left(
\begin{array}{ccc}
a_{1} & c_{1} & b_{3} \\
c_{1}^{\ast } & a_{2} & c_{2} \\
b_{3} & c_{2}^{\ast } & a_{3}%
\end{array}%
\right) ,\text{ }T_{2}=\left(
\begin{array}{cc}
a_{4} & b_{1} \\
b_{1}^{\ast } & a_{5}%
\end{array}%
\right) ,\text{ }T_{3}=\left(
\begin{array}{cc}
a_{6} & b_{2} \\
b_{2}^{\ast } & a_{7}%
\end{array}%
\right) .\text{ }  \label{eq48}
\end{equation}

After the PT, the density matrix $\rho _{ij}$ transfers to
\begin{equation}
\rho _{ij}^{T}=\text{diag}(T_{1}^{\prime },T_{2,}^{\prime }\text{ }%
T_{3}^{\prime },\text{ }a_{1},\text{ }a_{3}),  \label{eq49}
\end{equation}%
with
\begin{equation}
T_{1}^{\prime }=\left(
\begin{array}{ccc}
a_{8} & b_{1} & b_{3} \\
b_{1}^{\ast } & a_{2} & b_{2} \\
b_{3} & b_{2}^{\ast } & a_{9}%
\end{array}%
\right) ,\text{ }T_{2}^{\prime }=\left(
\begin{array}{cc}
a_{4} & c_{1} \\
c_{1}^{\ast } & a_{6}%
\end{array}%
\right) ,\text{ }T_{3}^{\prime }=\left(
\begin{array}{cc}
a_{5} & c_{2} \\
c_{2}^{\ast } & a_{7}%
\end{array}%
\right) .  \label{eq50}
\end{equation}%
Here, we have considered that the elements $a_{1}$ and $a_{3}$ are always
positive since they characterize the probability of finding two particles in
the states $|\uparrow \rangle |\downarrow \rangle $ and $|\downarrow \rangle
|0\rangle $, which have no relationship with entanglement. In order to
obtain the entanglement, we calculated the $17$ useful elements as follows
\begin{align}
a_{2} =&1-2\underset{k}{\sum }|C_{k,n_{1},n_{0},n_{-1}}|^{2}\frac{N-n_{0}}{N%
}+\frac{2}{N(N-1)}\underset{k}{\sum }|C_{k,n_{1},n_{0},n_{-1}}|^{2}%
\{n_{1}n_{-1}+\frac{1}{2}[n_{1}(n_{1}-1)+n_{-1}(n_{-1}-1)]\},  \notag
\label{eq51} \\
a_{3} =&\frac{1}{N(N-1)}\underset{k}{\sum }%
|C_{k,n_{1},n_{0},n_{-1}}|^{2}n_{1}n_{-1},  \notag \\
a_{4} =&a_{5}=\frac{1}{2}\underset{k}{\sum }|C_{k,n_{1},n_{0},n_{-1}}|^{2}%
\frac{N-n_{0}}{N}+\frac{M}{2N}-\frac{1}{N(N-1)}\underset{k}{\sum }%
|C_{k,n_{1},n_{0},n_{-1}}|^{2}[n_{1}n_{-1}+n_{1}(n_{1}-1)],  \notag \\
a_{6} =&a_{7}=\frac{1}{2}\underset{k}{\sum }|C_{k,n_{1},n_{0},n_{-1}}|^{2}%
\frac{N-n_{0}}{N}-\frac{M}{2N}-\frac{1}{N(N-1)}\underset{k}{\sum }%
|C_{k,n_{1},n_{0},n_{-1}}|^{2}[n_{1}n_{-1}+n_{-1}(n_{-1}-1)],  \notag \\
a_{8} =&\frac{1}{N(N-1)}\underset{k}{\sum }%
|C_{k,n_{1},n_{0},n_{-1}}|^{2}n_{1}(n_{1}-1),  \notag \\
a_{9} =&\frac{1}{N(N-1)}\underset{k}{\sum }%
|C_{k,n_{1},n_{0},n_{-1}}|^{2}n_{-1}(n_{-1}-1),  \notag \\
c_{1} =&c_{2}=\frac{1}{N(N-1)}\underset{k}{\sum }C_{k,n_{1},n_{0},n_{-1}}^{%
\ast }C_{k,n_{1}+1,n_{0}-2,n_{-1}+1}\sqrt{n_{0}(n_{0}-1)(n_{1}+1)(n_{-1}+1)},
\notag \\
b_{1} =&\frac{1}{N(N-1)}\underset{k}{\sum }%
|C_{k,n_{1},n_{0},n_{-1}}|^{2}n_{0}n_{1},  \notag \\
b_{2} =&\frac{1}{N(N-1)}\underset{k}{\sum }%
|C_{k,n_{1},n_{0},n_{-1}}|^{2}n_{0}n_{-1},  \notag \\
b_{3} =&\frac{1}{N(N-1)}\underset{k}{\sum }%
|C_{k,n_{1},n_{0},n_{-1}}|^{2}n_{1}n_{-1}.
\end{align}

\begin{figure}[tbh!]
\includegraphics[scale=0.5]{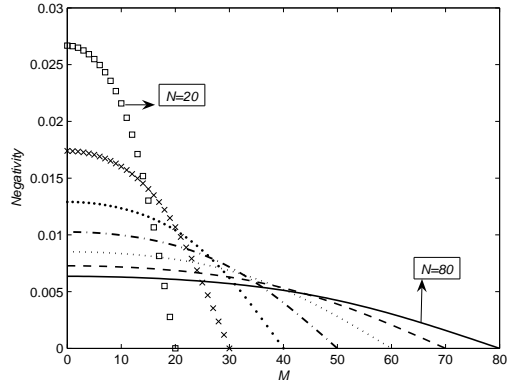}
\caption{The negativity of the Dicke states $|J,M\rangle $ with positive $M$%
. The number of particles from top to bottom is $20, 30,...,80$.}
\label{f1}
\end{figure}

\begin{figure}[tbh]
\includegraphics[scale=0.5]{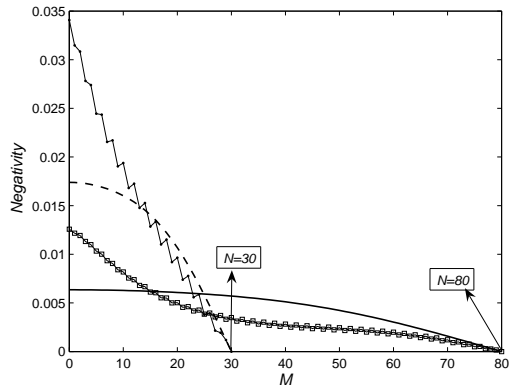}
\caption{Comparison of negativity between the Dicke states and the states
with equal probability combination of all states $|n_{1},n_{2},n_{3}\rangle $
in the subspace $M$. We take $N=30$ (on the left side) and $N=80$ (on the
right side). The solid line and dashed line correspond to the Dicke states.}
\label{f2}
\end{figure}

We can obtain the negative eigenvalues of three matrixes by solving the
eigenvalue equation. Substituting those eigenvalues into the formula of the
negativity, the entanglement can be calculated. Specially, for the case $M=0$%
, the negativity can be reduced to simpler form given by~\cite{Wang6}. The
properties of entanglement in the Dicke states with $20-80$ particles are
shown in Fig.~\ref{f1}. We observe that as $|M|$ decreases, the negativity
is a monotone increasing function, and comparing with others states, the
state $|J,0\rangle $ possesses the maximal entanglement. This property is
different from the concurrence of the Dicke states of multi-particle for the
spin-$1/2$ case. In addition, as $|M|$ increases, the maximal value of the
negativity decreases. Considering that those Dicke states $|J,M\rangle $
with $|M|<N-1$ are linear combination of different states $%
\{|n_{1},n_{0},n_{-1}\rangle \}$, the states $|J,M\rangle $ with $|M|<N-1$
actually forms a subspace. We can construct the states which are equal
probability combination of different states $|n_{1},n_{0},n_{-1}\rangle $ in
the subspace mentioned above. The negativities in these states are compared
with those of the Dicke states. The consequences show that, with the equal
probability combination, there are some advantages in the generation of
negativity, specially for the cases $M=0$. Here, we consider $N=30, 80$, and present
the negativities of two cases, as shown Fig.~\ref{f2}.

\section{Conclusions}

\label{sec:concl}

In summary, we have investigated the construction of Dicke states for high-spin particles based on that of Dicke states for the spin-$1/2$ case.
For three high-spin cases (spin-$1$, $3/2$ and $2$) with given particle numbers and spin magnetic quantum numbers,
the sets of constraint equations are found to determine all the basis states in the number representation as well as the corresponding normalized superposition coefficients, in terms of which the Dicke states are explicitly expressed in the number representation.
As a byproduct, we give a rule to construct all the anti-symmetric states in these high-spin systems
and show that the number of anti-symmetric states is rather limited.
Finally, in terms of the negativity, the entanglement properties for spin-$1$ cases including specific pure states
of two particles and the Dicke states of many particles are discussed.
Our results may contribute to the applications of high-spin systems in quantum information science
due to the crucial importance of Dicke states.

\section*{Acknowledgments}
We acknowledge fruitful discussions with Professor Yang Ming. This work is supported by the Natural Science Foundation of Anhui Province of China (Grant No.~1408085QA15), the National Natural Science Foundation of Special Theoretical Physics (Grant No.~11447174), the National Natural Science Foundation of China (Grant No.~11504140), the natural science foundation of Jiangsu province of China (Grant No. BK20140128) and the Fundamental Research Funds for the Central Universities(Grant No.~JUSRP51517).

\end{document}